\input harvmac.tex %\draftmode

\def\Asym#1#2{\vcenter{\vbox{\drawbox{#1}{#2}
              \kern-#2pt       % line up boxes
              \drawbox{#1}{#2}}}}

\Title{\vbox{\rightline{hep-th/9908163} \rightline{CERN-TH/99-255}
}}
{\vbox{\centerline{Superconformal Field Theories, Multiplet Shortening,}
\centerline{and the AdS$_5$/SCFT$_4$ Correspondence}}}

%\bigskip %\medskip

\centerline{Sergio Ferrara} \smallskip{\it
\centerline{CERN Geneva, Switzerland}}
\centerline{\tt
Sergio.Ferrara@cern.ch}

\vskip .3cm
\centerline{Alberto Zaffaroni} \smallskip{\it
\centerline{CERN Geneva, Switzerland}}
\centerline{\tt
Alberto.Zaffaroni@cern.ch}

\vskip .1in

%\vglue .3cm

\noindent
We review the unitarity bounds and the multiplet shortening of UIR's of
4 dimensional
superconformal algebras $SU(2,2|N)$, ($N=1,2,4$) in view of their dual role
in the AdS/SCFT correspondence. Some applications to KK spectra,
non-perturbative states and stringy states are given.
\vskip 2.5truecm
{\it To appear in the proceedings of the} Mosh\'e Flato {\it Conference, 
``Advances and Prospects in Physical Mathematics'', Dijon
(France), 5-8 September 1999.}

\noindent
CERN-TH/99-255
\Date{August 99}

\lref\M{ J. M. Maldacena,  Adv. Theor. Math. Phys. 2 (1998) 231,
hep-th/9705104.}
\lref\Wone{E. Witten, Adv. Theor. Math. Phys. 2 (1998) 253, hep-th/9802150.}
\lref\GKP{S. S. Gubser, I. R. Klebanov and A. M. Polyakov, Phys. Lett. B428
(1998) 105, hep-th/9802109.}
\lref\KRvN{H. J. Kim, L. J. Romans and P. van Nieuwenhuizen, Phys. Rev. D32
(1985) 389.}

\lref\GT{C. W. Gibbons and P. K. Townsend, Phys. Rev. Lett. 71 (1993) 3754}

%\lref\gun{M. Gunaydin and D. Minic, {\it Singletons, Doubletons and M-theory},
%%hep-th/9802047.}
\lref\MG{M. Gunaydin and N. Marcus, Class. Quantum Grav. 2 (1985) L11.}
%\lref\fztwo{S. Ferrara and B. Zumino, Nucl. Phys. B134 (1978) 301.}
\lref\GRWone{M. Gunaydin, L. J. Romans and N. P. Warner, Phys. Lett. 154B
(1985) 268}
%; M. Pernici, K. Pilch and P. van Nieuwenhuizen, Nucl. Phys. B259 (1985) 460.}
\lref\FFone{M. Flato and C. Fr\o nsdal, Lett. Math. Phys 2 (1978) 421; Phys.
Lett. 97B (1980) 236; J. Math. Phys. 22 (1981) 1100; Phys. Lett. B172 (1986)
412.}
\lref\FFtwo{M. Flato and C. Fr\o nsdal, Lett. Math. Phys. 8 91984) 159.}

\lref\AFGJ{
 D. Anselmi, D. Z. Freedman, M. T. Grisaru, A. A. Johansen, Phys. Lett. B394
(1997) 329, hep-th/9608125; Nucl. Phys. B256 (1998) 543, hep-th/970804.}
\lref\AEFJ{D. Anselmi, J. Erlich, D. Z. Freedman and  A. Johansen, Phys.Rev.
D57 (1998) 7570, hep-th/9711035.}
\lref\GUK{S. Gukov, Phys. Lett. B439 (1998) 2, hep-th/9806180.}
\lref\BERG{O. Bergman, Nucl. Phys. B525 (1998) 104, hep-th/9712211; O. Bergman
and B. Kol, Nucl. Phys. B536 (1998) 149, hep-th/9804160.}
\lref\FFthree{M. Flato and C. Fr\o nsdal, {\it Essays in Supersymmetries},
Math. Phys. Stud. 8, D. Reidel, Dordecht (1986).}
\lref\FMMR{D. Z. Freedman, S. D. Mathur, A. Matusis, L. Rastelli, Nucl. Phys. B
546 (1999) 96, hep-th/9804058.}
%\lref\tse{H. Liu and A. A. Tseytlin, {\it D=4 Super Yang Mills, D=5 gauged
%%supergravity and D=4 conformal supergravity}, hep-th/9804083.}
%\lref\seib{S. Lee, S. Minwalla, M. Rangamani and N. Seiberg, {\it Three-Point
%%Functions of Chiral Operators in D=4, N=4 SYM at Large N}, hep-th/9806074.}
\lref\FFZ{S. Ferrara, C. Fr\o nsdal and A. Zaffaroni, Nucl. Phys. B532 (1998)
153, hep-th/9802203.}
%\lref\grillo{S. Ferrara, A. F. Grillo and R. Gatto, Ann. Phys. 76 (1973) 161.}
%\lref\grillotwo{S. Ferrara, A. F. Grillo and R. Gatto, Lett. Nuovo Cimento 2
%%(1971) 1363. S. Ferrara, A. F. Grillo, R. Gatto and G. Parisi, Nucl. Phys.
%%B49 (1972) 77; Nuovo Cimento 19 (1974) 667.}
\lref\AFone{L. Andrianopoli and S. Ferrara, Phys. Lett. B430 (1998) 248,
hep-th/9803171; Lett. Math. Phys. 46 (1998) 26, hep-th/9807150.}
\lref\AFtwo{L. Andrianopoli and S. Ferrara, {\it On short and long SU(2,2/4)
multiplets in the AdS/CFT correspondence}, hep-th/9812067.}
%\lref\grillothree{S. Ferrara, A. F. Grillo, R. Gatto and G. Parisi, Nucl.
%%Phys. B49 (1972) 77; Nuovo Cimento 19 (1974) 667.}
\lref\B{B. Binegar, Phys. Rev. D34 (1986) 525.}
%\lref\fggp{S. Ferrara, R. Gatto and  A. F. Grillo, Springer Tracts in Modern
%%Physics, vol. 67 (Berlin-Heidelberg), Springer, New York,  (1973).}
%\lref\fggptwo{S. Ferrara, R. Gatto, A. F. Grillo and G. Parisi, in {\it Scale
%%and Conformal symmetry in hadron physics}, edited by R. Gatto, Wiley, New
%%York, (1983).}
\lref\mack{G. Mack, Comm. Math. Phys. 55 (1977) 1.}
%\lref\macktwo{G. Mack, J. Phys. 34, Colloque C-1 (Suppl. au no. 10) (1973)
%%79.}
%\lref\mackthree{G. Mack and I. Todorov, Phys. Rev. D8 (1973) 1764.}
%\lref\mack{G. Mack and A. Salam, Ann. Phys. 53 (1969) 174.}
%\lref\fe{S. Ferrara, Nucl. Phys. B77 (1974) 73.}
\lref\FGG{S. Ferrara, A. F. Grillo and R. Gatto, Phys. Rev. D9, (1974) 3564.}
\lref\GMZone{M. Gunaydin, D. Minic and  M. Zagermann, Nucl. Phys. B534 (1998)
96, hep-th/9806042.}
\lref\GMZtwo{M. Gunaydin, D. Minic and  M. Zagermann, Nucl. Phys. B544 (1999)
737, hep-th/9810226.}
%\lref\nic{D. Z. Freedman and H. Nicolai, Nucl. Phys. B237 (1984) 342.}
\lref\HW{P. S. Howe and  P. C. West,  Phys. Lett. B389 (1996) 273,
hep-th/9607060;
 Phys.Lett. B400 (1997) 307, hep-th/9611075 .}
\lref\GPPZ{L. Girardello, M. Petrini, M. Porrati and A. Zaffaroni, JHEP 9812
(1998) 022, hep-th/9810126.}

\lref\FGPW{D. Z. Freedman, S. S. Gubser, K. Pilch and N. P. Warner,
{\it Renormalization Group Flows from Holography--Supersymmetry and a
       c-Theorem}, hep-th/9904017.}
\lref\DP{V. K. Dobrev and V. B. Petkova, Phys. Lett. B162 (1985) 127; Lett.
Math. Phys. 9 91985) 287; Fortschr. Phys. 35 (1987) 7, 537.}
\lref\GIKO{A. Galperin, E.A. Ivanov, S. Kalitsyn and V. Ogievetsky, Class.
Quant. Grav. 1 (19984) 469; Class. Quant. Grav 2 (1985) 155.}
\lref\SKI{W. Skiba, {\it Correlators of Short Multi-Trace Operators in N=4
Supersymmetric Yang-Mills}, hep-th/9907088.}
\lref\BKRS{M. Bianchi, S. Kovacs, G. Rossi and  Y. S. Stanev, {\it On the
logarithmic behaviour in N=4 SYM theory}, hep-th/9906188.}
\lref\FZ{S. Ferrara and A. Zaffaroni, Phys. Lett. B431 (1998) 49,
hep-th/9803060; {\it Bulk Gauge Fields in AdS Supergravity and
Supersingletons}, hep-th/9807090.}
\lref\HST{P. Howe, K.S. Stelle and  P.K. Townsend, Nucl. Phys. B192 (1981)
332.}
\lref\BFH{B. Binegar, C. Fr\o nsdal and  W. Heidenreich, J. Math. Phys.24
(1983) 2828.}
\lref\O{H. Osborn, Annals Phys. 272 (1999) 243, hep-th/9808041.}
\lref\KT{ S. M. Kuzenko and  S. Theisen, {\it Correlation Functions of
Conserved Currents in N = 2 Superconformal
Theory}, hep-th/9907107.}
\lref\K{K. Konishi, Phys. lett B135 (1984) 439.}
\lref\FPZ{S. Ferrara, M. Porrati and  A. Zaffaroni, Lett. Math. Phys. 47 (1999)
255, hep-th/9810063.}
\lref\I{K. Intriligator, Nucl. Phys. B551 (1999) 575, hep-th/9811047;
K. Intriligator and W. Skiba, {\it Bonus Symmetry and the Operator Product
Expansion of N=4
       Super-Yang-Mills}, hep-th/9905020.}
\lref\HLS{R. Haag, J. T. Lopuszanski and M. Sohnius, Nucl. Phys. B88 (1975)
257.}
\lref\N{W. Nahm, Nucl. Phys. B135 (1978) 149.}
\lref\T{P. K. Townsend, {\it p-Brane Democracy}, hep-th/9507048.}
\lref\SS{{\it Supergravities in Diverse Dimensions}
A. Salam and E. Sezgin ed., North-Holland 1989.}
\lref\BG{ I. Bars and  M. Gunaydin, Commun. Math. Phys.91 (1983) 31; M.
Gunaydin, J. Math. Phys. 29 (1988) 1275.}
\lref\KW{I. R. Klebanov and  E. Witten, Nucl. Phys. B536 (1998) 199,
hep-th/9807080; {\it AdS/CFT Correspondence and Symmetry Breaking},
hep-th/9905104.}
\lref\G{S. S. Gubser, Phys. Rev. D59 (1999) 025006, hep-th/9807164;  S. S.
Gubser and I. R. Klebanov, Phys. Rev. D58 (1998) 125025, hep-th/9808075.}
\lref\R{L. J. Romans, Phys. Lett.153B (1985) 392.}
\lref\GRWtwo{M. Gunaydin, L.J. Romans and N.P. Warner, Nucl. Phys. B272 (1986)
598.}
\lref\Wtwo{E. Witten, JHEP 9807 (1998) 006, hep-th/9805112.}
\lref\CDDF{A. Ceresole, G. Dall'Agata, R. D'Auria and S.
       Ferrara, {\it
Spectrum of Type IIB Supergravity on $AdS_5 \times T^{11}$: Predictions on N
       = 1 SCFT's}, hep-th/9905226.}
\lref\JRD{D.  P. Jatkar and  S. Randjbar-Daemi, {\it Type IIB string theory on
$AdS_5 \times T^{nn'}$}, hep-th/9904187.}
\lref\CDD{A. Ceresole, G. Dall'Agata and R. D'Auria, {\it KK Spectroscopy of
Type IIB Supergravity on $AdS_5\times T^{11}$}, hep-th/9907216.}
\lref\C{G. Chalmers and K. Schalm, {\it Holographic Normal Ordering and
Multi-particle States in the AdS/CFT
       Correspondence}, hep-th/9901144.}
\lref\PARK{J. Park, Int. J. Mod. Phys. A13 (1998) 1743, hep-th/9703191;
{\it Superconformal Symmetry and Correlation Functions}, hep-th/9903230.}

\newsec{Introduction}
The classification of Super Poincar\'e and Super Anti-de-Sitter algebras
\refs{\HLS,\N,\T} in diverse dimensions has played a major role since the early
days of supersymmetry and supergravity.

Super Poincar\'e algebras have been widely used to classify
supersymmetric field theories and in constructing generic supergravity
theories in diverse dimensions and with different numbers of supersymmetries
\SS . Their central extension (more precisely the central extension
of the supertranslational algebras) had a major role with the inclusion
of p-branes supergravity solutions; a particular case is related to the black
hole classification based on the BPS properties of the solutions.

Supergravity solutions for BPS p-branes are considered to be limits of
analogous solutions in quantum theories of gravity, such as string theory
or M-theory.
However, in certain cases, due to non renormalization theorems of N-extended
supersymmetry, such solutions are exact or at least reliable approximations
of stringy or M-theory solutions.

The recent interplay between Anti-de-Sitter supergravity and conformal field
theories follows from the conjectured duality \M\ between the string (or
M-theory)
background  describing the AdS$_{p+2}$ horizon geometry \GT\ of black p-branes
and the conformal field theory living on their world-volume, which can be
thought
as the AdS-boundary. This correspondence claims that the {\it correlation
functions} of composite operators in the CFT can be computed as {\it bulk}
Green functions of AdS states of supergravity or string theory
\refs{\GKP,\Wone}.

In the present discussion, we only consider the case of D3-branes in type IIB
supergravity. The  world-volume theories are $N=1,2,4$ 4d superconformal
field theories \refs{\Wone,\FZ} and the corresponding bulk theory is type IIB
on
AdS$_5\times$X$_5$, for internal manifolds X$_5$ giving $N=2,4,8$
supergravities \refs{\GRWone,\KRvN,\R,\GRWtwo}, respectively.
Operators in the SCFT are associated with
the KK excitations, coming from the harmonic expansion on X$_5$.

These theories and their UIR's are therefore related to the $SU(2,2|N)$
algebras for $N=1,2,4$ \refs{\FFtwo,\BG,\GMZone,\GMZtwo}. The case $N=4$, as we
will see, deserves a special treatment since the $U(1)$ factor of the
$U(N)=U(1)\times SU(N)$ R-symmetry becomes an outer automorphism of the
superalgebra \refs{\DP,\B,\GMZtwo,\I} (this is a particular case of the
$SU(N|M)$ superalgebra with $N=M$).

This contribution reviews the unitary bounds and multiplet shortening
for highest weight UIR's of 4d superconformal algebras in view of the
AdS$_5$/CFT$_4$ correspondence \refs{\FZ,\FGPW,\CDDF}. Special emphasis will be
given to different properties of the $N=1,2,4$ cases, due to the different
supersingleton representations \refs{\FFone,\FFtwo,\DP,\B}. The latter are the
basic degrees of freedom of the superconformal theories on the boundary. In
this respect, the pioneering work of
Moshe Flato, in collaboration with Chistian Fr\o nsdal, has a crucial role in
this analysis. The above authors extended, in particular, the unitary bounds of
UIR's of $SU(2,2)$ to $SU(2,2|1)$, and classified the multiplet shortenings,
which just correspond to the {\it thresholds} of the unitary bounds \FFtwo.
Extension of this analysis came later \refs{\DP,\B} and is the latter that we
will use in the context of the AdS$_5$/CFT$_4$ correspondence.

In Sections II and III, the unitarity bounds for $N=1,2,4$ will be reviewed.
UIR's satisfying these bounds have an interpretation in terms of superconformal
boundary fields \refs{\Wone,\FZ,\AFone}. Some examples will be discussed.
More importantly, the case of multiplet shortening corresponds to CFT operators
of protected dimensions. This happens either if a boundary operator is a {\it
conserved current}, thus corresponding to a massless bulk field, or if it is a
shortened multiplet of the {\it chiral} type \refs{\Wone,\FFZ}. Intermediate
multiplets which are not chiral also exist that have protected dimensions
\refs{\FGPW,\CDDF}. They can be formally obtained by the product of a conserved
supercurrent with a {\it chiral} operator.

For extended superconformal algebras with $N=2,4$ a new (exceptional) series of
shortened
multiplets also exists which has no analogous in $N=1$. This has to do with the
fact that $N=2$ and $N=4$ algebras admit {\it self-conjugate} supersingletons
representations \refs{\GMZone,\DP,\B} (unlike the $N=1$ case).

In Section IV, the application of different type of shortened multiplets to
different boundary conformal field theories will be given. Shortened multiplets
allow to make a detailed comparison between the AdS$_5$ bulk theory and the
boundary superconformal field theory. Particular examples are the $N=4$ KK
towers
\refs{\KRvN,\MG,\Wone,\AFone}, corresponding to an exceptional series of the
$N=4$
superalgebra as well as multitrace operators of the $N=4$ Yang-Mills theory
\AFtwo, some of them having protected dimensions \refs{\AFtwo,\BKRS,\SKI}.
A richer structure exists for $N=1$ superconformal theories where shortened
multiplets with anomalous dimensions can appear with  precise associated
supergravity states. This is the case for X$_5=T^{1,1}=SU(2)\times SU(2)/U(1)$,
unique example \R\ of a smooth manifold where both the bulk and the boundary
theories have been worked out in full detail \refs{\KW,\G,\JRD,\CDDF,\CDD}.

\newsec{The unitary representations and shortening for $SU(2,2|N)$, $N=1,2$}

In this Section we consider the unitarity bounds for the highest-weight
representations of the $SU(2,2|N)$ superalgebra.

For the case of the $SU(2,2)$ algebra itself, a given UIR is denoted,
following Flato and Fr\o nsdal \refs{\FFone,\FFtwo,\FFthree}, as
$D(E_0,J_1,J_2)$
where $E_0,J_1,J_2$ are the quantum number of the highest-weight state \GMZone,
given by a finite UIR of the maximal compact subgroup $SU(2)\times SU(2)\times
U(1)$. The UIR's fall in three series \refs{\FGG,\M,\BFH},
\eqn\bound{\eqalign{a&)\, J_1J_2\ne 0\cr b&)\, J_2J_1=0\cr c&)\, J_1=J_2=0}
\qquad\qquad\qquad \eqalign{E_0&\ge 2+J_1+J_2\cr E_0&\ge 1+J\cr
E_0&=0}}
In the bulk interpretation, the inequality corresponds to massive
representations in AdS$_5$, the bound in a) corresponds to massless bulk
particles of {\it spin} $J_1+J_2$ while the bound in b)
corresponds to singletons of spin $J$.

Note that in the AdS/CFT correspondence the bulk-boundary quantum numbers
$(E_0,J_1,J_2)$ refer to the {\it compact} basis for the AdS states, while they
refer to the non-compact basis $SL(2,C)\times O(1,1)$ for the boundary
conformal operators \refs{\Wone,\GMZtwo}. The highest weight state in AdS
corresponds to
a conformal operator $O(x)$ at $x=0$, so the AdS energy $E_0$ corresponds to
the conformal dimension $\Delta_0$ and the $(J_1,J_2)$
quantum numbers correspond to the Lorentz spin of $O(x)$.

In the CFT, the bound a) corresponds to conformal conserved currents of spin
$J=J_1+J_2$,
\eqn\boundII{E_0=2+J_1+J_2\qquad
\partial^{\alpha_1\dot\alpha_1}J_{\alpha_1...\alpha_{2J_1},\dot\alpha_1...\dot\alpha_{2J_2}}(x)=0,\qquad J_1J_2\ne 0}
while the bound b) corresponds to massless spin $J$ conformal fields on the
boundary,
\eqn\boundIItwo{\eqalign{&E_0=1+J\cr &\qquad}\qquad\qquad
\eqalign{&\partial^{\alpha_1\dot\alpha_1}O_{\alpha_1...\alpha_{2J}}=0\cr
&\partial^2 O(x)=0}
\qquad\qquad \eqalign{&J\ne 0\cr
&J=0}}
The case c) corresponds to the identity representation.

Let us now consider the case of $SU(2,2|N)$ superalgebras \refs{\BG,\DP}.
In this case, the highest weight state is denoted by
$D(E_0,J_1,J_2;r,a_1,...,a_{N-1})$ where the quantum numbers in the bracket
denote a UIR of $SU(2,2)\times U(1)\times SU(N)$, $r$ being the quantum number
of the $U(1)$ R-symmetry
and $a_1,...,a_{N-1}$ the Dynkin labels of a UIR of the non-abelian symmetry
$SU(N)$. We will denote by R the $U(1)$ inside $U(N)$.

Note that for $N\ne 4$, the $SU(2,2|N)$ algebra is both
a subalgebra and a quotient algebra of $U(2,2|N)$, since
the supertrace generator (which is a central charge)
can be eliminated by a redefinition of the $U(1)$ generator R of $U(N)$
\refs{\DP,\B}. This redefinition is not however possible for $N=4$ since $R$
drops from
the supersymmetry anti-commutators and it becomes an outer automorphism of the
algebra
\refs{\B,\FPZ,\I}. In this case we have therefore two inequivalent algebras
(which do not include the $U(1)$ R generator),
$PSU(2,2|4)$ and $PU(2,2|4)$, depending whether $r=0$ or $r\ne 0$.

We will adopt the convention that $r$ is always the quantum number of the
$U(1)$ generator of $U(2,2|N)$, so it will be the $U(1)$ subgroup of $U(N)$
for $N=1,2$ while it will be a central $U(1)$ for $N=4$ \DP.

In the boundary CFT language, UIR's can be realized as conformal superfields.
The superhighest weight state corresponds to
a superfield $\phi(x,\theta)$ at $x=\theta=0$ \refs{\DP,\AFone,\FZ,\PARK}.

The unitarity bounds for $SU(2,2|1)$ were found by Flato and Fr\o nsdal
\FFtwo. They generalize the cases a),b) and c) of eq. \bound . They are,
\eqn\superbounda{a )\qquad E_0\ge 2+2J_2+r\ge 2+2J_1-r\qquad (\hbox{or}\,
J_1\rightarrow J_2,\, r\rightarrow -r)\qquad J_1,J_2\ge 0}
which implies
\eqn\cond{E_0\ge 2+J_1+J_2,\qquad r\ge J_1-J_2,\qquad 2+2J_1-E_0\le r\le
E_0-2-2J_2}
\eqn\superboundb{b )\,\,\,\,\,\,\,\,\qquad E_0=r\ge 2+2J-r\qquad
(J_2=0,\,J_1=J,\, \hbox{or} \, J_1=0,\,J_2=J,\, r\rightarrow -r)}
which implies $E_0\ge 1+J$,
\eqn\superboundc{c )\,\,\,\,\,\,\,\,\qquad\qquad\qquad\qquad\qquad\qquad
E_0=J_1=J_2=r=0\qquad\qquad\qquad\qquad\qquad}
which corresponds to the identity representation.

Shortening in the case a) corresponds to
\eqn\shorta{E_0=2+2J_2+r,\qquad (r\ge J_1-J_2)\qquad\qquad (\hbox{or}\,
J_1\rightarrow J_2,\, r \rightarrow -r)}
This is a semi-long AdS$_5$ multiplet \FGPW\ or, in conformal language, a {\it
semiconserved} superfield \refs{\O, \CDDF},
\eqn\op{\bar
D^{\dot\alpha_1}L_{\alpha_1...\alpha_{2J_1},\dot\alpha_1...\dot\alpha_{2J_2}}(x,\theta,\bar\theta)=0,\qquad (\bar D^2L_{\alpha_1...\alpha_{2J_1}}=0\, \hbox{for}\, J_2=0)}
(in our conventions $\theta$ carries $\Delta=-1/2,r=1,\bar\theta$ has
$\Delta=-1/2,r=-1$).

Maximal shortening in case a) corresponds to $E_0=2+J_1+J_2,\, r=J_1-J_2$.
This is a conserved superfield which satisfies both left and right constraints:
\eqn\cons{ \bar
D^{\dot\alpha_1}J_{\alpha_1...\alpha_{2J_1},\dot\alpha_1...\dot\alpha_{2J_2}} =
D^{\alpha_1}J_{\alpha_1...\alpha_{2J_1},\dot\alpha_1...\dot\alpha_{2J_2}}=0}

The shortening in b) corresponds to {\it chiral superfields}.
Maximal shortening in b) to {\it massless} chiral superfields, i.e. chiral
singleton representations: $E_0=r=1+J$. The corresponding superfield, for
$E_0=r$ satisfies,
\eqn\chiral{\bar D^{\dot\alpha}S_{\alpha_1...\alpha_{2J}}=0}
and, for $E_0=1+J$, it also satisfies
\eqn\chiraltwo{D^{\alpha_1}S_{\alpha_1...\alpha_{2J}}=0\qquad (D^2 S=0,\,
\hbox{for}\, J=0)}
These equations are the supersymmetric version of \boundII\ and \boundIItwo.

We may call, with an abuse of language, off-shell singletons chiral superfields
in the sense that in an interacting conformal field theory singletons may
acquire anomalous dimension and fall in \chiral.

It is also evident, from superfield multiplication, that by suitable
multiplication of several free supersingletons one may get any other superfield
of type \op, \cons\ or \chiral.

Note that superfields obeying \op,\chiral\ may have anomalous dimensions since
the shortening condition just implies a relation between $E_0$ and $r$ without
fixing their value.

The basic singleton multiplets for $N=1$ gauge theories correspond to $J=0,1/2$
in \chiral, i.e. chiral scalar superfields $S$ (Wess-Zumino multiplets)
and Yang-Mills field strength multiplets $W_\alpha$. Any other conformal
operator is obtained by suitable multiplication of these two sets of basic
superfields.

In type IIB on $T^{1,1}$ long, semi-long and chiral multiplets do indeed occur
\refs{\KW,\G,\CDDF}. Chiral WZ singleton multiplets have in this case an
anomalous
dimension $\gamma=-1/4$ ($\Delta=1+\gamma$) and R-symmetry $R=3/4$.

If we adopt the concept of {\it bulk} masslessness as corresponding to an UIR
that is contained in the product of two supersingletons there are other
massless bulk representations which are obtained by chiral multiplication of
two singleton
representations,
\eqn\mul{\eqalign{D(1+J_1,J_1,0;1+J_1)\otimes
D(1+J_2,J_2,0;1+J_2)&=D(2+J_1+J_2,J,0;2+J_1+J_2)\cr &|J_1-J_2|\le J\le
J_1+J_2}}
These representations do not occur in the AdS/SCFT correspondence unless
$J_1=0,J_2=0,1/2$ since they do not correspond to global symmetries of the
SCFT.
One can indeed show that the symmetry of $N=1$ SYM theory, as in the dual of
type IIB on $AdS_5\times T^{1,1}$, do not allow such UIR's.

We now turn to the $SU(2,2|2)$ superalgebra. In this case
the highest weight state is $D(E_0,J_1,J_2;r,l)$ where $l$ is the spin label of
the R-symmetry
$SU(2)$. The series of UIR's are now,
\eqn\superboundal{\alpha )\qquad E_0\ge 2+2J_2+r+2l\ge 2+2J_1-r+2l\,\,
(\hbox{or}\, J_1\rightarrow J_2,\,r\rightarrow -r,\,l\rightarrow l)\qquad
J_1,J_2\ge 0}
which implies
\eqn\cond{E_0\ge 2+J_1+J_2+2l,\,\, r\ge J_1-J_2,\,\, 2+2J_1+2l-E_0\le r\le
E_0-2-2J_2-2l}
\eqn\superboundbe{\beta )\,\,\qquad\qquad E_0=r+2l\ge 2+2J-r+2l\qquad
(J_2=0,J_1=J,\, or \, J_1=0,J_2=J,r\rightarrow -r)}
which implies $r\ge 1+J,\,E_0\ge 1+J+2l$,
\eqn\superboundga{\gamma )\,\,\,\,\,\,\,\,\,\,\,\qquad\qquad\qquad\qquad
 E_0=2l,\qquad\qquad\qquad\qquad\qquad\qquad J_1=J_2=r=0\qquad\,\,\,\,}

The new phenomenon in $N=2$ is that
supersingletons appear in two different series, $\beta$) and $\gamma$).
This is related to the fact that
the $\gamma$) series does not reduce to the identity representation only,
as for $N=1$, but contains both massless bulk ($l=1$) and
massless boundary (supersingletons) ($l=1/2$) representations.
Let us discuss in detail the cases of massless bulk and massless boundary
representations.

Supersingletons in $\beta$) correspond to the shortening condition
\eqn\shorttwo{E_0=r=1+J,\qquad l=0}
This is the highest weight state $D(1+J,J,0;1+J,0)$.
The highest spin in these multiplets is $J+1$. For $J=0$ we get the Yang-Mills
multiplet.

The $l=1/2$ supersingleton in $\gamma$)  is the hypermultiplet
($D(1,0,0;0,1/2)$).

Since supermassless bulk representations are defined as the product of two
supersingletons we may obtain bulk massless representations by multiplying
either two supersingletons (in a chiral and anti-chiral manner) in $\beta$) or
by
a supersingleton in $\beta$) and the $l=1/2$ supersingleton in $\gamma$), or
two $l=1/2$
supersingletons in $\gamma$).

The spin 2 massless bulk states are obtained
by chiral anti-chiral multiplication of the $J=0$ chiral supersingleton in
$\beta$), giving the shortened representations
in $\alpha$) with $J_1=J_2=r=l=0$. This is the massless graviton multiplet:
$D(2,0,0;0,0)$.

The massless spin 1 multiplet is obtained by multiplying two hypermultiplets
and corresponds to the $l=1$ UIR of case $\gamma$), giving $D(2,0,0;0,1)$.

Finally, the massless spin 3/2 multiplet is obtained by multiplying the $J=0$
supersingleton in $\beta$) with the hypermultiplet, giving a bulk massless
multiplet
corresponding to the shortening $r=1,J=0,l=1/2$ in $\beta$), i.e.
$D(2,0,0;1,1/2)$. The superconformal realization of these superfields as
current multiplets was given in \refs{\HST,\AFGJ,\KT}.

Obviously these examples can be extended by replacing the chiral
vector multiplet by an arbitrary spin $J$ supersingleton. Also one can get, as
for $N=1$, {\it chiral} bulk massless multiplets by chiral multiplication of
two spin $J$ supersingletons as in \mul.

In the AdS$_5$/CFT$_4$ correspondence the relevant supersingletons are the {\it
chiral} vector multiplets (in $\beta$) for $J=0$) and the hypermultiplets (in
$\gamma$)
for $l=1/2$).
Multiplet shortening in the AdS/CFT correspondence will be further discussed in
Section IV. Here we note that the series $\gamma$) contains short multiplets
with $E_0=2l$ for each value of $l$. These states can be explicitly constructed
by multiplying $2l$ hypermultiplet singletons and their dimension is not
renormalized and coincide with the canonical one. On the other hand, short
multiplets in $\beta$), constructed using also chiral vector singletons, may
have anomalous dimension since, due to \superboundbe, $E_0$ is related to the
arbitrarily
valued $U(1)$ charge $r$ by $E_0=r+2l$.

\newsec{UIR's of $PSU(2,2|4)$ and $PU(2,2|4)$ and shortening conditions.}
The $N=4$ superalgebra is of great interest because it corresponds to $N=4$
superconformal Yang-Mills theory. In the dual description, it lives
at the boundary of AdS$_5$ \refs{\M,\GKP,\Wone}. The supergravity theory
emerges as the low
energy limit of type IIB string theory compactified on AdS$_5\times$X$_5$,
where X$_5$ is a manifold preserving $N=8$ such as $S_5$ or
$RP_5$ \refs{\KRvN,\Wtwo}.

The UIR's classes of the previous Section enter in this case as follows.

Let us consider a highest weight representation $D(E_0,J_1,J_2;r,p,k,q)$,
where $r$ is the central charge eigenvalue and $(p,k,q)$ are the $SU(4)$
Dynkin labels. Then the three unitary series show up as follow \DP:
\eqn\superboundA{\eqalign{A)\qquad E_0\ge 2+2J_2+r+{1\over 2}(p+2k+3q)&\ge
2+2J_1-r+{1\over 2}(3p+2k+q)\cr
&(\hbox{or $J_1\rightarrow J_2$, $r\rightarrow -r$, $(p,k,q)\rightarrow
(q,k,p)$})}}
which implies
\eqn\conseqA{\eqalign{&E_0\ge 2+J_1+J_2+p+k+q,\qquad r\ge {1\over 2}(p-q)
+J_1-J_2\cr &2+2J_1+{1\over 2}(3p+2k+q)-E_0\le r\le E_0-2-2J_2-{1\over
2}(p+2k+3q)}}
Shortening occurs when,
\eqn\shorten{E_0=2+2J_2+r+{1\over 2}(p+2k+3q)}
and maximal shortening when
\eqn\maxshortenA{E_0=2+J_1+J_2+p+k+q,\qquad r={1\over 2}(p-q) +J_1-J_2}
Generic massless bulk multiplets correspond to $p=k=q=0$. They correspond to
table 12 in \GMZone.

When we have a strict inequality in \superboundA, we obtain generic
massive multiplets with
highest spin $(J_1+2,J_2+2)$ in the $(p,k,q)$ representation of $SU(4)$.
Note that all these long multiplets (with $J_{MAX}\ge 4$) must necessarily
be associated to stringy states, since in supergravity all the states have
$J\le 2$.
\eqn\superboundB{\eqalign{B)\qquad E_0= r+{1\over 2}(p+&2k+3q)\ge
2+2J-r+{1\over 2}(3p+2k+q)\cr
&(\hbox{for $J_2=0$,$J_1=J$ or $J_1\rightarrow J_2$, $r\rightarrow -r$,
$(p,k,q)\rightarrow (q,k,p)$})}}
which implies
\eqn\conseqB{E_0\ge 1+J+p+k+q\qquad r\ge 1+J+{1\over 2}(p-q)}
The maximal shortening occurs when
\eqn\maxshorteningB{E_0=1+J+p+k+q}
and the chiral supersingletons
occur for $p=k=q=0$. They correspond to table 6 of \GMZone. When a strict
inequality occurs in \conseqB\ and $p=k=q=0$, we have
UIR's described by $N=4$ chiral (left-handed) superfields with highest spin
$(J+2,0)$ in $SU(4)$ singlets with $\Delta=r-4$.
\eqn\superboundC{C)\qquad\qquad\qquad E_0=p+k+q,\qquad\qquad\qquad r={1\over
2}(p-q),\qquad\qquad\qquad J_1=J_2=0}
which gives supersingleton representations for $p\, (\hbox{or $q$})=1,k=0$ or
$p=q=0,k=1$. The self-conjugate (Yang-Mills) supersingleton with spin 1
corresponds
to the $D(1,0,0;0,0,1,0)$ highest weight and the spin 3/2 supersingleton to
$D(1,0,0;1/2,1,0,0)$. They correspond to table 1 and 2 of \GMZtwo.

The UIR's of the $PSU(2,2|4)$ superalgebra are obtained by setting $r=0$ in the
previous shortening conditions. Correspondingly, we also abolish the entry
corresponding to $r$ in the symbol $D$ denoting highest weight states. We get
the three classes of UIR's,
\eqn\superboundAprime{A')\qquad\qquad\,\,\,\,\,\, E_0\ge
2+J_1+J_2+p+k+q,\qquad\qquad\qquad J_2-J_1\ge
{1\over 2}(p-q)\qquad}
Maximal shortening occurs when,
\eqn\maxshortenAprime{E_0=2+J_1+J_2+p+k+q,\qquad J_2-J_1={1\over 2}(p-q)}
Massless bulk multiplets correspond to $p=k=q=0$ and $J_1=J_2$.
\eqn\superboundBprime{\eqalign{B')\qquad\qquad\,\,\,\,\,\, E_0&= {1\over
2}(p+2k+3q)\ge 2+2J+{1\over 2}(3p+2k+q)\qquad\qquad\qquad\cr
&(\hbox{$J_2=0$,$J_1=J$ or $J_1\rightarrow J_2$, $(p,k,q)\rightarrow
(q,k,p)$})}}
with
\eqn\conseqBprime{E_0\ge 1+J+p+k+q\qquad 1+J\le {1\over 2}(q-p)}
Maximal shortening occurs when $1+J= {1\over 2}(q-p)$,
with highest weight $D(3+3J+2p+k,J,0;p,k,p+2+2J)$. No supersingletons appear in
this series.
\eqn\superboundCprime{C')\qquad\qquad\qquad\qquad E_0=2p+k,\qquad\qquad\qquad
p=q,\qquad\qquad\qquad J_1=J_2=0}
The highest weight states are $D(2p+k,0,0;p,k,p)$. The $p=0, k\ge 2$ UIR's
correspond
to the KK states of type IIB on AdS$_5\times S_5$, the $k=2$ case being
associated with the bulk graviton multiplet.
The $p=0,k=1$ UIR corresponds to the only supersingleton of the $PSU(2,2|4)$
algebra \refs{\B,\GMZone}. The infinite sequence of UIR's with $p=0$,
multiplets with $J_{MAX}=2$, have been obtained in \MG\ with the oscillator
construction. They correspond to the harmonic \GIKO\ holomorphic superfields of
\HW.
The case $p\ne 0$ may be relevant for multiparticles supergravity states, as
will be discussed in the next Section.

To summarize the structure of UIR's of $SU(2,2|N)$ algebras: generic massive
supermultiplets, occurring when a strict inequality in
\superbounda ),\superboundal ),\superboundA ) holds,
have multiplicity $2^{4N}$, generic multiplets when the inequality in
\superboundb ),\superboundbe ),\superboundB ) occurs have multiplicity $2^{2N}$
(the same is true for the series c), $\gamma$), C)).
At the threshold of the unitarity bounds, these multiplicities shrink. For
example, massless bulk multiplets have multiplicity $2^{2N}$, while massless
boundary multiplets (supersingletons) $2^N$.

\newsec{Applications to the AdS/CFT correspondence}
The previous analysis of shortening conditions finds applications in the
AdS/CFT correspondence, where many shortened multiplets are realized.

For $N=1$ theories, the prototype is type IIB on AdS$_5\times T^{1,1}$,
where all type of shortening a), b) and c) have been shown to occur for
$J_1,J_2\le 1/2$ \refs{\G,\CDDF,\JRD}.

For $N=2$ theories also all types of shortening occur. Shortening $\beta$)
includes, for instance, $N=2$ tensor multiplet KK recurrences, while shortening
$\gamma$) corresponds to $N=2$ vector multiplet KK recurrences \refs{\GUK,\KW}.

The $N=4$ case includes, for the $PSU(2,2|4)$ algebra, all the KK states of
type IIB on AdS$_5\times S^5$ \refs{\KRvN,\MG,\Wone,\FFZ,\AFone}. It also
contain other states corresponding to multiparticle supergravity states, which
can in principle be analyzed \AFtwo; they correspond to multitrace conformal
operators in the Yang-Mills theory \refs{\FMMR,\C,\BKRS,\SKI}.

Shortening in A$'$) typically occurs in free-field theory but not in the
interacting $N=4$ theory. For instance, the $D(E_0,0,0;0,0,0)$ highest weight
corresponds to the Konishi multiplet, which undergoes a shortening in a free
theory ($E_0=2$) while is long in the interacting Yang-Mills theory
since $E_0> 2$ \refs{\K,\AFGJ,\AEFJ,\FZ}. In the AdS/CFT correspondence, it is
indeed the simplest example  of operator corresponding to a stringy state.
Another example of operator which is short only in free theory is the highest
weight $D(3,0,0;0,0,2)$ in B$'$), which contains the superpotential of the
$N=4$
theory when it is written in $N=1$ notations. It is not a primary conformal
superfield in the interacting Yang-Mills theory since it can be obtained by
total antisymmetrization of the product of three self-conjugate supersingletons
\refs{\FZ,\GPPZ}.

The sequence C$'$) has an interesting application to multitrace operators,
since it predicts that primary operators with highest-weight
$D(2p+k,0,0;p,k,p)$
are not renormalized.
In a single trace operator, representations with $p\ne 0$ are not primary
since they involve at least a partial antisymmetrization of the Yang-Mills
supersingletons, which makes it a descendent via equations of motion
\refs{\Wone,\FFZ,\AFone}. The same argument does not apply to multitrace
operators.
Consider the simplest case. Out of two single-trace primary operators
 in the $20_R$ ($p=0,k=2$) we can construct by multiplication several
multitrace operators.
They are in the symmetric product $(20_R\times 20_R)_S=105+84+20_R+1$.
The previous discussion implies
that the $105$ and $84$ ($(0,4,0)$ and $(2,0,2)$) can fit into a shortening
condition and are
not renormalized, so that anomalous dimensions can only show up for the $20_R$
and singlet pieces.
 This result has been recently confirmed in \refs{\AFtwo,\BKRS,\SKI} for the
$(0,4,0)$ piece and in \BKRS\ for the $84$ piece, by an explicit perturbative
computation.

It is interesting to speculate at this point whether UIR's of the $PU(2,2|4)$
algebra with $r\ne 0$
can occur.
It is obvious that, since the Yang-Mills theory is built in terms of the
self-conjugate multiplet with $r=0$, any local operator constructed out of such
multiplet will have $r=0$ and will be in some representation of $PSU(2,2|4)$.
However, as speculated in \GMZtwo, if some AdS$_5$ states are dual to dyonic
non-perturbative states of the Yang-Mills theory, then necessarily
a new sector with $r\ne 0$ will appear at the non-perturbative level.
Natural candidates are the 1/4 BPS dyonic states of the Yang-Mills theory
\BERG,
which would correspond to the second exceptional supersingleton representation
in C) (with $J_{MAX}=3/2$).
To strengthen this interpretation, it would be interesting
 to understand how the central $U(1)$ acts on spacetime type IIB fields.
%It is easy to analize discrete subgroup of this $U(1)$.
%The center $Z_4$ of $SU(4)$, which has an obvious
%action on the Yang-Mills fields, is contained in the $U(1)$
%group. The $Z_4$ subgroup of the type IIB $U_Y(1)$ symmetry,
%whose S-duality like
%action on the Yang-Mills fields can be found, for example, in \refs{\FPZ,\I},
% on the other hand, is not a subgroup of $U(1)$, since it does not commute
%with the supersymmetries. However,
%the action of the two $Z_4$ on the Yang-Mills fields is identical,
%modulo a change of sign of all the fields \I.
%This common $Z_2$ subgroup of $U_Y(1)$ and $U(1)$ acts on the type
%IIB fields by changing sign to the NS-NS and R-R antisymmetric tensor
%fields.
We could conjecture that non-perturbative states with
$r\ne 0$ are obtained by considering type IIB $(p,q)$ five-branes and strings.
The mentioned 1/4 BPS states of the Yang-Mills theory
certainly fall into this category, since they are obtained as junctions of
$(p,q)$ strings \BERG.
The central charge of the $U(2,2|4)$ algebra is an
 $SU(4)$ singlet and may temptatively be interpreted as coming from a
wrapped five-brane, in some generalization of the construction in  \Wtwo.
The question whether such states are actually BPS saturated in AdS$_5$ is
somewhat unclear.

\centerline{\bf Acknowledgements}
S. F. is supported in part by the DOE under grant DE-FG03-91ER40662, Task C,
the NSF grant PHY94-07194, and by the ECC Science
Program SCI*-CI92-0789 (INFN-Frascati). We would like to thank M. Bianchi and
D.Z. Freedman for helpful discussions.

\listrefs
\end